\documentclass[]{interact}

\usepackage{xcolor}
\usepackage{soul}

\usepackage{moreverb}
\usepackage{cancel}
\usepackage{hyperref} 

\usepackage{epstopdf}
\usepackage[caption=false]{subfig}

\usepackage[longnamesfirst,sort]{natbib}
\bibpunct[, ]{(}{)}{;}{a}{,}{,}
\renewcommand\bibfont{\fontsize{10}{12}\selectfont}

\theoremstyle{plain}

\theoremstyle{definition}

\theoremstyle{remark}

\usepackage{setspace}
\begin{document}

\articletype{Quantitative methods}

\title{Spatial Risk Patterns-ANOVA: Multivariate Analysis Of Suicide-Related Emergency Calls}

\author{
\name{P. Escobar-Hernandez\textsuperscript{a,b}\thanks{P. Escobar-Hernandez. Email: pablo.escobar@uv.es}, A. López-Quílez\textsuperscript{a}, F. Palmí-Perales\textsuperscript{a}, M. Marco\textsuperscript{b}}
    \affil{\textsuperscript{a}Department of Statistics and Operational Research, Universitat de València, Spain; \textsuperscript{b}Department of Social Psychology, Universitat de València, Spain}
}

\maketitle

\begin{abstract}
 Multivariate spatial disease mapping has become a pivotal part of everyday practice in social epidemiology. Despite the existence of several specifications for the relation between different outcomes, there is still a need for a new strategy that focuses on comparing the spatial risk patterns of different subgroups of the population. This paper introduces a new approach for detecting differences in spatial risk patterns between different populations at risk, using suicide-related emergency calls to study suicide risks in the Valencian Community (Spain).
\end{abstract}


\section{Introduction}

\hspace{0.3cm} The analysis of social problems across spatial and temporal contexts has emerged as an essential aspect of social epidemiology. These models, originally intended for disease mapping \citep{SpatEp2016}, have proven effective in mapping various social issues, such as suicide, intimate partner violence, or child maltreatment. Moreover, it is a common practice to include sociodemographic covariates,  known as ecological regression. \citep{Kempf2005EncyclopediaOS}. \\

In practical applications, small geographic regions are often selected, creating challenges due to sparse data with minimal observations per area. In these cases, it is assumed that nearby areas are more similar than distant ones, which helps to smooth the associated risk for each area. Bayesian hierarchical models are useful in modeling the risk surface, leveraging strength from neighboring regions to achieve local smoothness through spatial random effects. These models are formulated using Gaussian Markov Random Fields \citep[GMRF]{Rue05}, typically considering neighbors areas that share a common boundary. \\

When multiple variables, outcomes or groups are recorded in each area, the resulting dataset transforms into a multivariate dataset. For example, different studies have been published analyzing the spatial overlap of intimate partner violence against woman (IPVAW) and child maltreatment \citep{Gracia18}, crimes against women \citep{Vicente2021MultivariateBS}, vehicle theft, larceny, and burglary \citep{Chung2019CrimeRM}, disadvantage variables \citep{Quick2021TheSS}, police calls reporting street-level violence and behind-closed-doors crime \citep{Marco21} and substantiated and unsubstantiated child maltreatment referrals \citep{Marco20}. \\

There are two main categories of multivariate models depending on the nature of the relation between the different outcomes: models based on multivariate conditional auto-regressive (MCAR) distributions \citep{Gelfand2003, Carlin2002}, a generalization of the conditional auto-regressive (CAR) distribution, and spatial factor analysis models \citep{Wang03}. \\

In general multivariate models, such as the MCAR, the possible relations between the different groups of interest are included in the model using the precision matrix. For instance, the MCAR distribution can be viewed as a conditionally specified probability model for interactions between space and a given characteristic of interest. Thus, MCAR takes into account dependence across space and between groups/diseases. However, practical difficulties arise from MCAR's complex dependence structure: most interaction's effects would be weakly identified by the data, so the dependence structure could not be properly identified. Strong prior distributions may improve identifiability but most of the time there is no genuine prior information \citep{Zhang10}. \\

On the contrary, in spatial factor analysis the different outcomes are considered to have one or more underlying, unobserved common spatial factors, that are estimated and combined using weights to determine the geographical pattern of a given outcome. A special case of spatial factor models are shared component models, where the spatial pattern for each different outcome is usually a combination of a shared common spatial effect and some individual specific spatial effects \citep{Held2005ASF, PalmiPerales2022BayesianIF}. \\

Expanding on the idea of shared spatial effects and increasing in complexity there is a series of different formulations \citep{Nobile00, Gelman02, Hodges07, Zhang10, MarDellOlmo14}, under the name of Smoothed Analysis-Of-Variance (SANOVA). SANOVA selects for effects that are large, removing most of those that are small \citep{Zhang10}. In contrast with MCAR models, that rely on estimating weakly identifiable parameters, SANOVA focuses on smoothing interactions, thus providing more stable and reliable results. \\

Among this formulation, \citet{Zhang10} considers series of orthogonal contrasts between diseases, i. e., a set of linear combinations of the outcomes we wish to model. Subsequently, \cite{MarDellOlmo14} proposed a reformulation that relies on using spatially structured random effects for each of the contrasts instead of employing such effects for the independent modeling of every one of the groups.  Since the linear combinations are defined by the user manually, the spatial dependence assumed is restricted to the combinations chosen. This is a reasonable assumption in most cases, since often the spatial pattern is not specific to a given group but rather due to the distribution of some risk factors which may influence more than one group. In consequence, we can attribute spatial dependence to certain combinations of the risk for some groups. Moreover, this formulation allows for extensions such as multivariate ecological regression or spatio-temporal models.\\

Lastly, \citet{Botella15} and \cite{MartinezBeneito17} extended the idea of the SANOVA model using a multidimensional approach, known as M-Model. The proposal is based on using linear combinations of proper CAR spatial effects, adjusting different loadings for each group. The M-Model takes its name from the $M$ matrix that defines the aforementioned loadings of the different CAR spatial effects for each outcome. \\

The differences between multidimensional and SANOVA have been discussed extensively. \citet{MartinezBeneito17} addressed the main issues with SANOVA in comparison to multidimensional approaches, such as the M-Model. In particular, multidimensional modeling does not need a specific selection of contrasts for making geographical comparisons, eliminating the contrast-dependency found in SANOVA results. In addition, SANOVA would have to include several interaction terms between contrasts to achieve the same flexibility as multidimensional models, thus losing the advantage of being less parameterized. \\

However, in some particular datasets, where data may not be sufficiently informative, the numerical approximations can fail to properly estimate the rich covariance structure of multidimensional approaches. In addition, the interpretation of the $M$ matrix that defines the loadings for each different CAR spatial effect for each outcome is not straightforward, hindering the conclusions that can be extracted regarding spatial-risk patterns of the different groups of interest. Finally, even if the model is perfectly specified and data is informative enough, the computational cost of using this type of models increases exponentially the number of spatial units. This implies that, although theoretically the number of outcomes that can be jointly analyzed is unlimited, in practice the computational cost of extending the model to a large geographical area increases significantly. \\

After reviewing the current state of the art, we considered that there was a place for an ANOVA-like specification, somewhere between the original Smoothed-ANOVA \citep{MarDellOlmo14} and his final generalization of the M-Model \citep{MartinezBeneito17}. This new modelization, designated as Spatial Risk Patterns-ANOVA (SRP-ANOVA), aims to determine whether different stratum of a given population presents different spatial risk patterns for a given outcome. Specifically, we will focus on a 2-way ANOVA, where each group has 2 different categories, adding up to 4 possible different groups. \\

The remainder of the paper is structured as follows. In Section 2 we present a detailed description of the new methodology proposed, named Spatial Risk Patterns-ANOVA (SRP-ANOVA). Section 3 contains the specifications of the implementation from a bayesian approach. Section 4 includes the analysis of spatial risk patterns in suicide, using suicide related-emergency calls as outcome, including 3 different groups (COVID-19 period, gender and type of caller) combined in pairs. Finally, Section 5 contains the discussion and conclusions extracted from this paper.

\section{Spatial Risk Patterns-ANOVA (SRP-ANOVA)}

\hspace{0.3cm} Let $E^i_{g}$ be the number of persons at risk in area $i$ ($i = 1,...,I$) and group $g$ ($g = 1,...,G$), with $Y^i_{g}$ corresponding to the number of cases observed in area $i$ in group $g$. $\theta^i_{g}$ represents the underlying true-area-group specific relative risk, as long as we can assume that the observations are conditionally independent random sample from a given probability distribution from the exponential family, typically a Poisson distribution:

\begin{equation}
    Y^i_{g} \sim \operatorname{Poisson}\left(E^i_{g} \theta^i_{g}\right).\\
\end{equation}

When estimating the expected values, we need to address the difference between multivariate and multidimensional analysis. In multivariate analysis, we are interested in studying several outcomes jointly. For example, child maltreatment and suicide rates, or two independent diseases like lung cancer and liver cancer. Therefore, $E^i_{g}$ is calculated individually for each outcome, i.e., the expected number of cases for a given outcome in a given area depends on the population of the area and the total number of cases for that given outcome. \\

In contrast, when we are using multidimensional analysis with an ANOVA-like design, we focus on the incidence of a given disease in different subgroups of the population. This is the case, for example, when we are studying the incidence of a given disease in male vs female population, or using different age groups. In this case, $E^i_{g}$ is calculated combining groups because the outcome is the same for each group and, therefore, we expect all groups to have the same risk if there is no difference among them. In consequence, for a given group in a given area, the number of expected cases depends on the sum of all the cases in all the groups and the population of the area divided by the number of groups. \\

However, it is important to note that whenever we use groups that are not equally represented in the general population, we should use the actual population at risk for each group. For example, if we are interested in a particular outcome and the difference between the local and foreign populations, the expected values must take into account the differences in population between the two groups. \\

The underlying risk $\theta^i_{g}$ is modeled in a log-scale, taking advantage of the capability of INLA for running models with several likelihoods \citep{Virgilio2020}. In particular, the log-relative risk for each group is treated as an independent model, but we are able to estimate a given spatial effect for one of the groups and then add it as a fixed copy to the rest of the groups. This allows us to sequentially add different shared spatial effects, that are estimated using a given group as reference and then built-in the rest of the groups. All the spatial effects included follow a Besag's specification with flat uniform prior distributions as priors assigned to the standard deviation parameters \citep{PalmiPerales2022BayesianIF}. \\

Specifically, we propose a sequential modelization strategy, where we increase the number of spatial effects on each step and determine if the fitting of the model has improved enough to justify the increase in complexity using DIC \citep{Spiegelhalter2002BayesianMO}. Since DIC penalizes excessive complexity, the inclusion of a spatial effect that does not significantly modify the previous shared spatial effects will be punished, giving a worse DIC score. This allows us to use DIC values to detect which spatial effects are different enough from zero and, consequently, which factors are responsible for some of the spatial differences observed within the given outcome of interest. Another possible option would be using WAIC  \citep{Watanabe2010AsymptoticEO} as a comparison metric, but since WAIC has being reported to have issues with correlated data \citep{Gelman2013UnderstandingPI}, such as spatial or temporal data, we have opted for using DIC. Nevertheless, we present the WAIC as well in Section 3. \\

The sequential steps are the following:

\begin{itemize}
    \item $M_{0}$: the outcome of interest does not exhibits a spatial risk pattern. This initial model includes an individual intercept for each given group that accounts for the 4 different baseline risk levels $\alpha_{1-4}$, as well as individual unstructured effects $\omega^i_{1-4}$ representing the heterogeneity effects for each area in each group:

        \begin{equation}
            \begin{aligned}
                & \log \left(\theta^i_{1}\right)=\alpha_1+\omega^i_1\\
                & \log \left(\theta^i_{2}\right)=\alpha_2+\omega^i_2\\
                & \log \left(\theta^i_{3}\right)=\alpha_3+\omega^i_3\\
                & \log \left(\theta^i_{4}\right)=\alpha_4+\omega^i_4
            \end{aligned}
        \end{equation}

    \item $M_{1}$: this models contemplates the possibility that each group presents spatial patterns different enough to justify using one for each group. Essentially, since there is no shared effect, this model is an independent univariate specification for each group. Spatial structured effects $\phi^i_{1-4}$ represent the individual spatial effects for each group, spatial unstructured effects $\omega^i_{1-4}$ represent the heterogeneity effects for each group and $\alpha_{1-4}$ are the intercepts that define the baseline risk level for each group: 

        \begin{equation}
            \begin{aligned}
                & \log \left(\theta^i_{1}\right)=\alpha_1+\omega^i_1+\phi^i_1\\
                & \log \left(\theta^i_{2}\right)=\alpha_2+\omega^i_2+\phi^i_2\\
                & \log \left(\theta^i_{3}\right)=\alpha_3+\omega^i_3+\phi^i_3\\
                & \log \left(\theta^i_{4}\right)=\alpha_4+\omega^i_4+\phi^i_4
            \end{aligned}
        \end{equation}
        
    \item $M_{2}$: this model is the first one to include shared spatial effects. The notation hereafter for the shared spatial effects is the following: $\phi^i_{jk}$ represents the spatial effect estimated using level $j$ of the first factor considered, and level $k$ of the second factor considered. The groups are included in the following order: $g_1$ corresponds to $\theta^i_{11}$, $g_2$ corresponds to $\theta^i_{12}$, $g_3$ corresponds to $\theta^i_{21}$ and $g_4$ corresponds to $\theta^i_{22}$.  In this particular model, the outcome of interest presents a spatial risk pattern $\phi^i_{11}$, estimated using the first level of the first factor and the first level of the second factor. Moreover, each group has a different baseline level, represented by $\alpha_{1-4}$, and an individual unstructured term $\omega^i_{1-4}$:

        \begin{equation}
            \begin{aligned}
                & \log \left(\theta^i_{1}\right)=\alpha_1+\omega^i_1+\phi^i_{11}\\
                & \log \left(\theta^i_{2}\right)=\alpha_2+\omega^i_2+\phi^i_{11}\\
                & \log \left(\theta^i_{3}\right)=\alpha_3+\omega^i_3+\phi^i_{11}\\
                & \log \left(\theta^i_{4}\right)=\alpha_4+\omega^i_4+\phi^i_{11}
            \end{aligned}
        \end{equation}
        
    \item $M_{3}$: the outcome of interest exhibits a spatial risk pattern $\phi^i_{11}$ estimated using the first level of the first factor and the first level of the second factor, and this pattern is shared for the rest of the groups. Moreover, the first factor exhibits certain particularities that modify the general spatial risk and is, therefore, shared by those two groups, in this case represented by $\phi^i_{21}$. In this case, $\phi^i_{21}$ is estimated using the second level of the first factor, and the first level of the second factor. As usual, the 4 different groups present a baseline level represented by $\alpha_{1-4}$ and the unstructured term $\omega^i_{1-4}$:

        \begin{equation}
            \begin{aligned}
                & \log \left(\theta^i_{1}\right)=\alpha_1+\omega^i_1+\phi^i_{11}\\
                & \log \left(\theta^i_{2}\right)=\alpha_2+\omega^i_2+\phi^i_{11}\\
                & \log \left(\theta^i_{3}\right)=\alpha_3+\omega^i_3+\phi^i_{11}+\phi^i_{21}\\
                & \log \left(\theta^i_{4}\right)=\alpha_4+\omega^i_4+\phi^i_{11}+\phi^i_{21}
            \end{aligned}
        \end{equation}
    
    \item $M_{4}$:  the outcome of interest presents a spatial risk pattern $\phi^i_{11}$ estimated using the first level of the first factor and the first level of the second factor, that is shared for the rest of the groups. Moreover, the second factor exhibits certain particularities that modify the general spatial risk and is, therefore, shared by those two groups, in this case represented by $\phi^i_{12}$. In this case, $\phi^i_{12}$ is estimated using the first level of the first factor, and the second level of the second factor. As usual, the 4 different groups present a baseline level represented by $\alpha_{1-4}$ and the unstructured term $\omega^i_{1-4}$:

        \begin{equation}
            \begin{aligned}
                & \log \left(\theta^i_{1}\right)=\alpha_1+\omega^i_1+\phi^i_{11}\\
                & \log \left(\theta^i_{2}\right)=\alpha_2+\omega^i_2+\phi^i_{11}+\phi^i_{12}\\
                & \log \left(\theta^i_{3}\right)=\alpha_3+\omega^i_3+\phi^i_{11}\\
                & \log \left(\theta^i_{4}\right)=\alpha_4+\omega^i_4+\phi^i_{11}+\phi^i_{12}
            \end{aligned}
        \end{equation}
    
    \item $M_{5}$: the outcome of interest presents a spatial risk pattern $\phi^i_{11}$ estimated using the first level of the first factor and the first level of the second factor. Moreover, the first factor exhibits certain particularities that modify the general spatial risk and is, therefore, shared by those two groups, in this case represented by $\phi^i_{21}$.  In addition, the second factor presents different spatial risk factors that are not reflected by neither the general share spatial effect nor the first factor spatial effect but do not depend on the level of the first factor. In this case, this is represented again with $\phi^i_{12}$. Finally, the 4 different groups present a baseline level represented by $\alpha_{1-4}$ and the unstructured term $\omega^i_{1-4}$:
    
        \begin{equation}
            \begin{aligned}
                & \log \left(\theta^i_{1}\right)=\alpha_1+\omega^i_1+\phi^i_{11}\\
                & \log \left(\theta^i_{2}\right)=\alpha_2+\omega^i_2+\phi^i_{11}+\phi^i_{12}\\
                & \log \left(\theta^i_{3}\right)=\alpha_3+\omega^i_3+\phi^i_{11}+\phi^i_{21}\\
                & \log \left(\theta^i_{4}\right)=\alpha_4+\omega^i_4+\phi^i_{11}+\phi^i_{21}+\phi^i_{12}
            \end{aligned}
        \end{equation}    
    
    \item $M_{6}$: this final model implies that there is an overall share spatial risk pattern for all the groups, the first factor presents an specific underlying risk factor that modifies the share common factor and there is an interaction between the first and the second factor such that, for each level of the first factor exists a different underlying spatial pattern for the second factor. In terms of formulation, this translates to:

        \begin{equation}
            \begin{aligned}
                & \log \left(\theta^i_{1}\right)=\alpha_1+\omega^i_1+\phi^i_{11}\\
                & \log \left(\theta^i_{2}\right)=\alpha_2+\omega^i_2+\phi^i_{11}+\phi^i_{21}\\
                & \log \left(\theta^i_{3}\right)=\alpha_3+\omega^i_3+\phi^i_{11}+\phi^i_{12}\\
                & \log \left(\theta^i_{4}\right)=\alpha_4+\omega^i_4+\phi^i_{11}+\phi^i_{12}+\phi^i_{22}
            \end{aligned}
        \end{equation}

    where:
    
        \begin{itemize}
            \item $\alpha_{1-4}$ are the 4 different intercepts that represent the baseline risk level for each group.
            \item $\omega^i_{1-4}$ are the 4 different unstructured heterogeneity spatial effects.
            \item $\phi^i_{11}$ captures an overall-shared spatial effect, estimated using the first level of the first factor and the first level of the second factor.
            \item $\phi^i_{21}$ captures the possible spatial variation explained by the second level of the first factor.
            \item $\phi^i_{12}$ captures the possible difference in the spatial risk pattern caused by the interaction of the first level of the first factor, and the second level of the second factor.
            \item $\phi^i_{22}$ captures the possible difference in the spatial risk pattern caused by the interaction of the second level of the first factor, and the second level of the second factor.
        \end{itemize}

\end{itemize}

It is straightforward to notice that, in several of the models, the order of the groups could have an impact on the estimation of the spatial effects. In particular, $M_0$ and $M_1$ can not be affected since essentially the specification is equivalent to an univariate scenario. In $M_2$ the reference group could produce different spatial patterns, which would affect the performance of the model. Therefore, it is reasonable to test the 4 different models where the reference group is exchanged.\\

In $M_3$ and $M_4$ we need to test if depending on the reference group for the given factor studied the overall fitting of the model is affected. This implies that we need to consider two different possibilities for each one of them. \\

$M_5$ can be viewed as a combination of models $M_3$ and $M_4$. Therefore, we need to consider the two possibilities of each specification depending on the reference level for each factor. Since we are working with 2-level factors, this adds up to 4 different models.\\

Finally, $M_6$ contemplates the interaction between the two factors. In this case, in addition to the 4 possibilities considered in $M_5$, we must also take into account that, depending on the order of inclusion of the factors, the estimation of the spatial effects will vary. This is cause by the nature of the interaction: the first factor is used for the spatial effects $\phi_{11}$ and $\phi_{12}$, which are shared by different groups, whereas the individual spatial effects that arise from the interaction among the two factors are not shared by other groups. In consequence, if the factor that is considered initially varies, the decomposition of the spatial variance may be equivalent, but the results are not. Consequently, we need to consider both possibilities. All of this adds-up to 8 different possibilities that need to be tested. A general overview of the spatial effects included and the number of combinations can be found in Table \ref{table1}.  \\

\begin{table}[!htbp]
\tbl{Total number of combinations tested for each model.}
        {\begin{tabular}{lcccccc} \toprule
         \textbf{Model} & \textbf{Spatial Structured Effects} & \textbf{Combinations of groups} \\
        \midrule
        $M_0:$ Intercepts & - & 1 \\
        $M_1:$ Individual spatial effects & $\phi_1$, $\phi_2$, $\phi_3$, $\phi_4$ & 1 \\
        $M_2:$ Shared Effect & $\phi_{11}$ & 4 \\
        $M_3:$ F1L2 Effect & $\phi_{11}$, $\phi_{21}$ & 2 \\
        $M_4:$ F2L2 Effect & $\phi_{11}$, $\phi_{12}$ & 2 \\
        $M_5:$ F1L2 Effect + F2L2 Effect & $\phi_{11}$, $\phi_{12}$, $\phi_{21}$ & 2 x 2 \\
        $M_6:$ F1L2 Effect * F2L2 Effect & $\phi_{11}$, $\phi_{12}$, $\phi_{21}$, $\phi_{22}$ & 2 x 2 x 2 \\
        \midrule
        \textbf{Total Number} & & 22 \\ \bottomrule
        \end{tabular}}
    \label{table1}
\end{table}

The principle behind this strategy is simple, with a similar approach to an ordinary ANOVA design, giving the researcher a tool to effectively compare spatial risk patterns for different groups and select the model that better fits the actual existing differences between the subgroups of interest. Furthermore, there is no initial supposition about the relationship between the different groups and, since the underlying structure is relatively simple, we are capable of testing all the options within a sensible amount of time. \\

\section{Bayesian Inference with INLA}

\hspace{0.3cm} This modelization is implemented within a Bayesian approach through the Integrated Nested Laplace Approximation (INLA) \citep{Rue09}. INLA aims to approximate the posterior distribution, in contrast with asymptotically exacts methods methods like the Markov Chain Monte Carlo (MCMC) methods, using a combination of Laplace approximations and numerical integration. It provides a fast and accurate way to estimate the posterior mean, variance, and other parameters without the need for extensive simulation. This makes INLA particularly attractive for complex hierarchical models, such as those encountered in disease mapping, spatial statistics, and other fields where efficient computation is essential. \\

In particular, we designed a custom function that estimates all the aforementioned models simultaneously. To ensure that the different spatial effects are identifiable and comparable, each random spatial effect is estimated with a sum-to-zero constraint and scaled to have an average variance of 1. \\

Different specifications for the random spatial effects could be implemented, which may affect the results obtained. Specifically, we have opted for using ICAR specifications, choosing flat uniform distributions for the standard deviation as prior's distributions \citep{PalmiPerales2022BayesianIF}.

\section{Analysis of Suicide-Related Emergency-Calls}

\hspace{0.3cm} Suicide poses a significant social and public health challenge, resulting in over 700,000 deaths globally each year \citep{WHO21}. Official data is showing an increasing trend in the last decades and there is a rising concern about the impact of COVID-19, \citep{GarcaFernndez2023DramaticIO} and the subsequent restrictive policy measures adopted on mental health \citep{WHO22}. In the pursuit of a more profound comprehension of suicide and the risk factors associated with it, epidemiological research has become an increasingly popular field of study \citep{Zalsman17}. Most studies have traditionally analyzed spatial and spatio-temporal patterns aggregated for the entire population but in recent years several studies have analyzed the spatial patterns of male and female suicide outcomes \citep{Chang11, Cayuela20, Lin19, Marco2024TheSD}. \\

In order to determine whether distinct risk groups exhibit disparate spatial risk patterns, we will employ the SRP-ANOVA strategy, as outlined in the preceding section. Our response variable will be suicide-related emergency calls in the Valencian Community (Spain), spanning the period from 2018 to 2023, distributed in 542 municipalities. In particular, data will be divided per period (preCOVID-19 vs postCOVID-19), gender of the person in crisis (male vs female) and type of person making the call (person in crisis vs bystander). The aforementioned factors will be combined in pairs, with the objective of studying the spatial overlap for each of the three combinations of categories (gender vs period, caller vs period and gender vs caller). All the modelization has been performed using an ASUS TUF Gaming F15 laptop, with a 11th Gen Intel Core i7 2.30GHz processor and 16GB of RAM. \\

Specifically, the distribution of calls for each combination of groups is the following:

\begin{itemize}
    \item Gender vs Period: our dataset presents 8788 calls from males preCOVID-19, 10908 calls from females preCOVID-19, 17189 calls from males postCOVID-19 and 22686 calls from females postCOVID-19. From the number of calls is straightforward to detect that some groups present significantly higher number of calls, which will translate in relative risk maps where most municipalities present higher values. \\
    \item Caller vs Period: our dataset presents 4062 calls from person in crisis preCOVID-19, 15921 calls from bystander preCOVID-19, 8177 calls from person in crisis postCOVID-19 and 31969 calls from bystander postCOVID-19. In this case the groups are more unbalanced, which suggests that the relative risks maps should present more acute differences. \\
    \item Gender vs Caller: for this set of groups our dataset contains 5554 calls from male person in crisis, 6322 calls from female person in crisis, 20152 calls from male bystanders and 27126 calls from female bystanders. In this case person in crisis and bystanders present the greatest differences, so we expect male and female groups to present similar relative risk maps for both person in crisis and bystanders. 
\end{itemize}

\subsection{Gender VS Period}

\hspace{0.3cm} Firstly, we have analyzed if the spatial risk pattern of male and female persons in crisis changed with COVID-19 and the measures adopted. In order to balance the groups we have used the same time span before and after COVID-19. Furthermore, it has been assumed that the underlying population at risk is identical for all four groups, given that the discrepancy between the male and female populations prior to and following the onset of the pandemic is minimal. \\

The best 5 models in terms of DIC are presented in Table \ref{table2}. For this combination of factors, 4 of them correspond to $M_5$. In particular, the best model is a $M_5$ specification using preCOVID-19 and Male as basis levels. The order of the factors in this case is not relevant. \\

\begin{figure}[!htbp]
    \centering
    \includegraphics[scale=0.28]{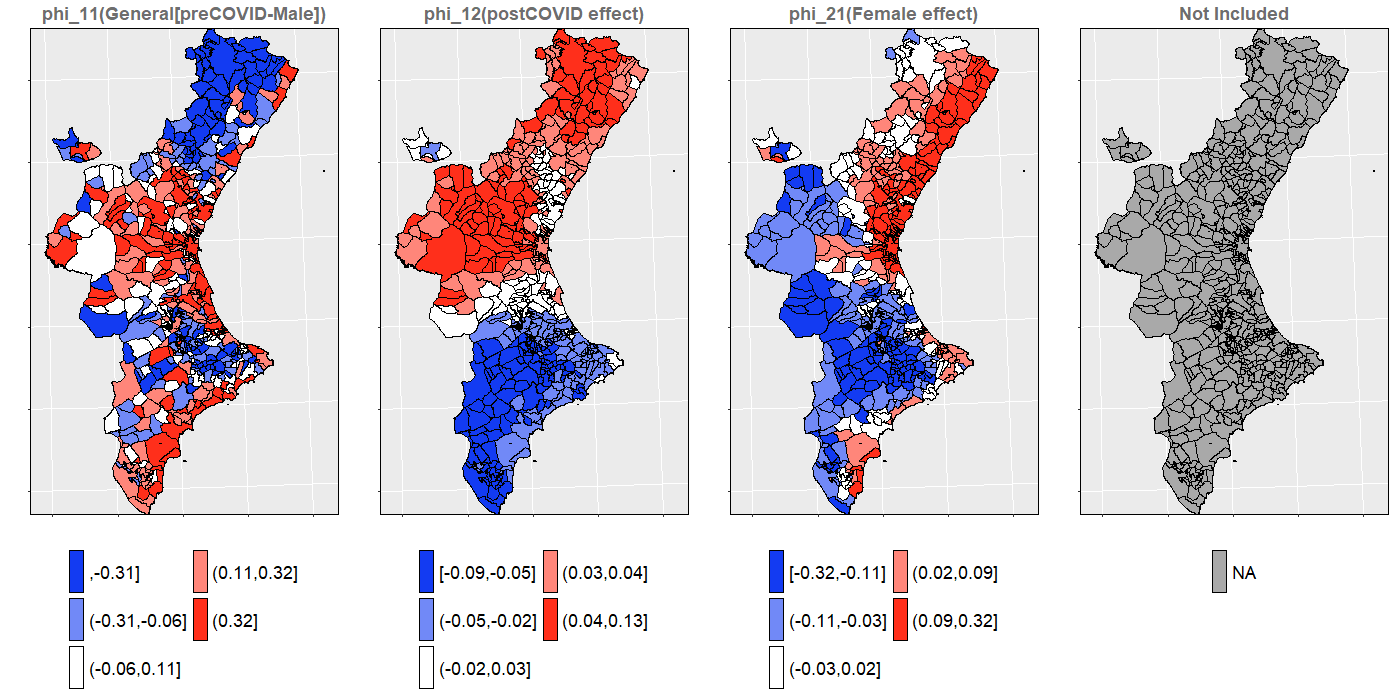}
    \caption{Spatial Effects of the best model for Gender VS Period in suicide-related emergency calls.}
    \label{fig:1}
\end{figure}

Spatial effects for the best model are represented in Figure \ref{fig:1}. The overall shared effect oscillates between -0.31 and 0.32 and presents a low risk cluster of municipalities in the north of the region, and some more in the inland south. The middle of the region and the coastline contain most of the areas with greater risks. 

The postCOVID-19 effect, oscillating between -0.09 and 0.13, implies that COVID-19 produced an increase in the risk for areas in the north and a decrease for areas in the south, with two clear clusters. Nevertheless, the size of this effect is considerably smaller. \\

Finally, the Female effect ranges between -0.32 and 0.32, suggesting that in this case the effect of the gender is greater than the period effect. In particular, the spatial effect demonstrates a distinct concentration of areas with negative values along the central and northern coastline, as well as a significant aggregation of municipalities with elevated risk in the central and southern inland regions.  \\

\begin{table}[!htbp]
\tbl{Top 5 Gender VS Period models in terms of DIC.}
        {\begin{tabular}{lcccc} \toprule
         \textbf{Model} & \textbf{Combination} & \textbf{DICs} & \textbf{WAICs} & \textbf{CPU (sec)} \\
        \midrule
        $M_5$ & preCOVID-19 + Male & 9131.4 & 9166.8 &  8.17 \\
        $M_5$ & postCOVID-19 + Male & 9131.6 & 9171.1 & 7.91 \\
        $M_5$ & preCOVID-19 + Female & 9132.5 & 9167.1 & 8.55 \\
        $M_6$ & Female * preCOVID-19 & 9132.6 & 9166.7 & 10.54 \\
        $M_5$ & postCOVID-19 + Female & 9133.1 & 9171.6 & 8.37 \\
        \bottomrule
        \end{tabular}}
    \label{table2}
\end{table}

\begin{figure}[!htbp]
    \centering
    \includegraphics[scale=0.28]{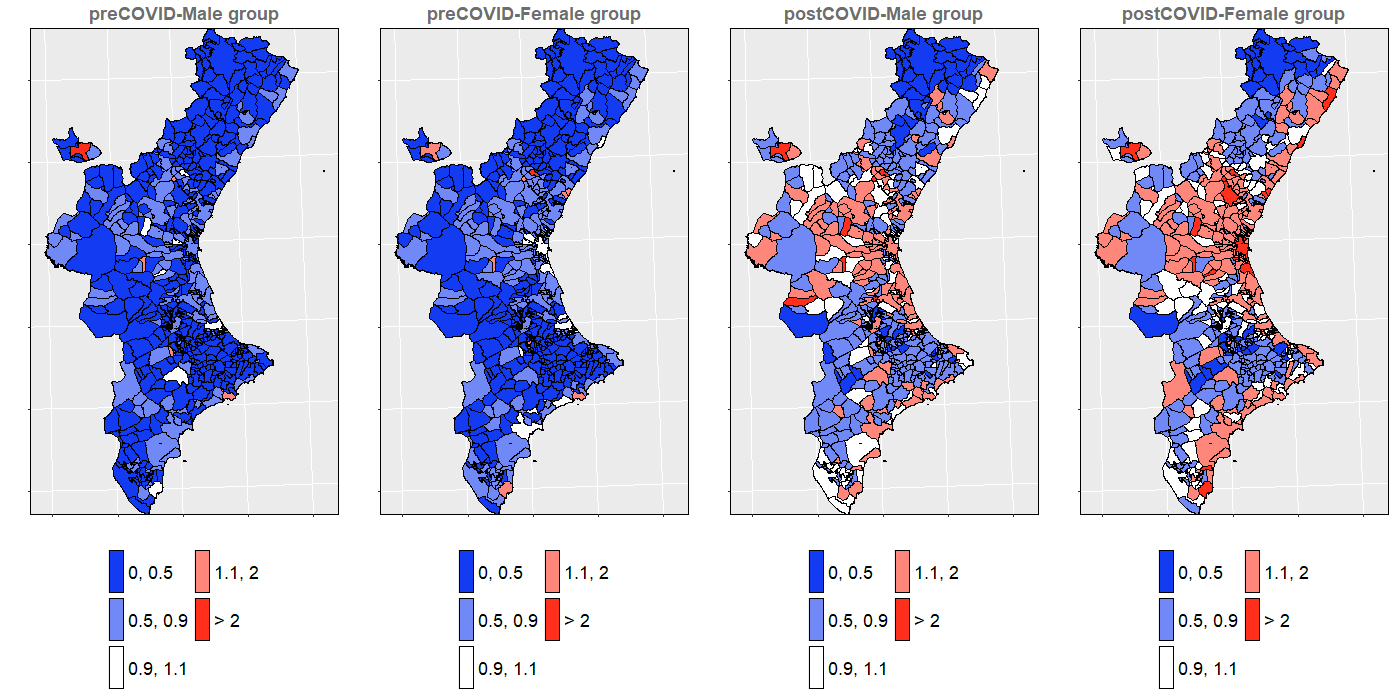}
    \caption{Relative Risks adjusted for the best model of Gender VS Period in suicide-related emergency calls.}
    \label{fig:2}
\end{figure}

Figure \ref{fig:2} illustrates the adjusted relative risks. In this case, the group with the highest risks is the female population postCOVID-19, with several high-risk municipalities located in the inland middle of the region. Furthermore, the distribution of high-risk areas is similar for males postCOVID-19. Finally, both the preCOVID-19 groups present mostly low-risk areas, with the exception of individual municipalities.

\subsection{Caller VS Period}

\hspace{0.2cm} Secondly, we will test if the spatial risk pattern differ taking into account the type of caller (person in crisis or bystander), pre and postCOVID-19.  Just like in the previous section, the groups present the same time span before and after COVID-19. In addition, since a given person could be both a bystander or a person in crisis, the population at risk remains the same for all the groups. \\

The best 5 models in terms of DIC are presented in Table \ref{table3}. For this combination of factors, 4 of them correspond to $M_6$. Specifically, the best model is a $M_6$ specification using person in crisis and preCOVID-19 as basis levels. In this case, the order of the factors is relevant, and for the best model factor 1 is caller and factor 2 is COVID-19. \\

\begin{table}[!htbp]
\tbl{Top 5 Caller VS Period models in terms of DIC.}
        {\begin{tabular}{lcccc} \toprule
         \textbf{Model} & \textbf{Combination} & \textbf{DICs} & \textbf{WAICs} & \textbf{CPU (sec)} \\
        \midrule
        $M_6$ & bystander * postCOVID-19 & 8571.7 & 8593.1 & 7.89 \\
        $M_6$ & preCOVID-19 * bystander & 8575 & 8587.6 & 7.81 \\
        $M_6$ & postCOVID-19 * bystander & 8575.1 & 8590.9 & 8.06 \\
        $M_6$ & bystander * preCOVID-19 & 8594.8 & 8628 & 6.34 \\
        $M_5$ & preCOVID-19 + bystander & 8360.2 & 8648.7 & 7.09 \\
        \bottomrule
        \end{tabular}}
    \label{table3}
\end{table}

\begin{figure}[!htbp]
    \centering
    \includegraphics[scale=0.28]{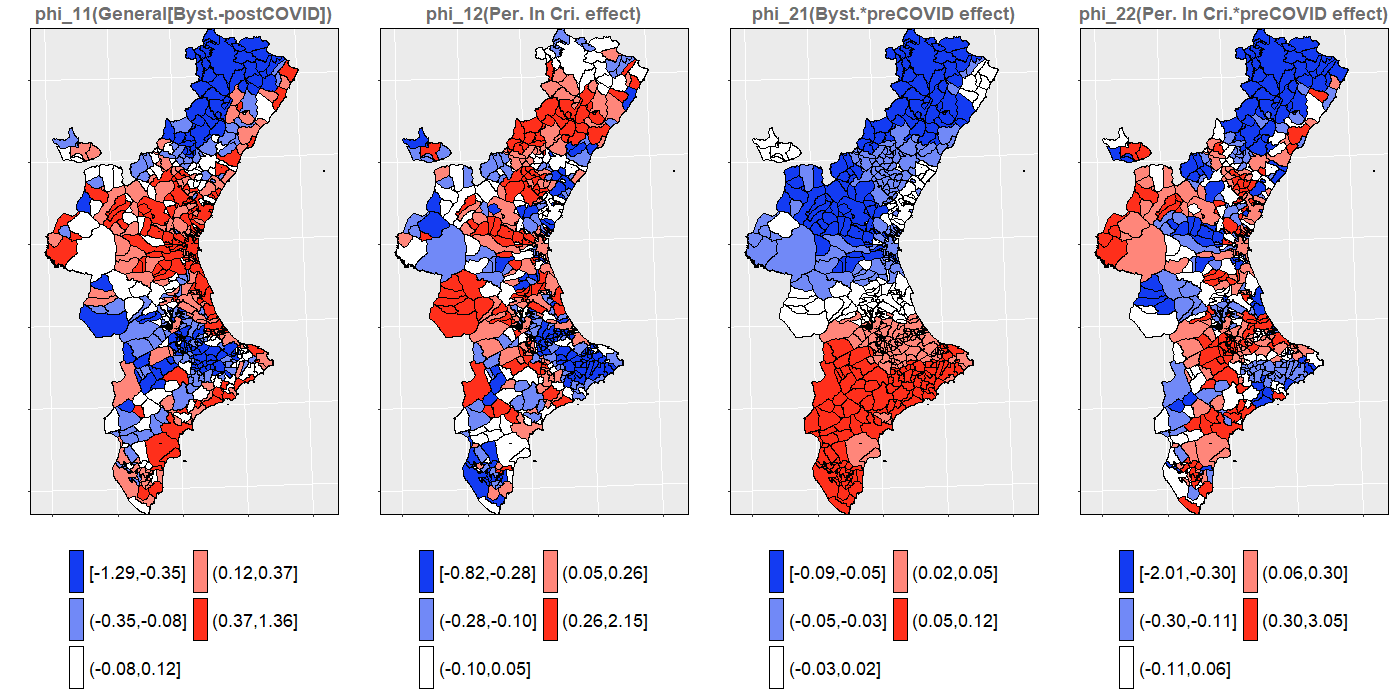}
    \caption{Spatial Effects of the best model for Caller VS Period in suicide-related emergency calls.}
    \label{fig:3}
\end{figure}

Spatial effects for the best model are represented in Figure \ref{fig:3}. The overall shared effect ranges between -1.29 and 1.36, presenting a clear cluster of higher risk areas in the center of the region and specially closer to the coastline on the east. Moreover, there are two clusters north and south inland with negative values for the spatial effects. 

On the contrary, the spatial effects for person in crisis show positive values in the north and south inland, predominantly rural areas. In this case, the size of the effect is respectable, oscillating between -0.82 and 2.15, suggesting the presence of municipalities where the risk for person in crisis is considerably higher. \\

Bystander preCOVID-19 interaction effect presents an extremely clear north-south distribution, with smaller values that slightly modify the previous spatial patterns, ranging between -0.09 and -0.12. Finally, the person in crisis preCOVID-19 interaction effect exacerbates the patterns observed in the general effect, presenting the most extreme values ranging between -2.01 and 3.05. \\

\begin{figure}[!htbp]
    \centering
    \includegraphics[scale=0.28]{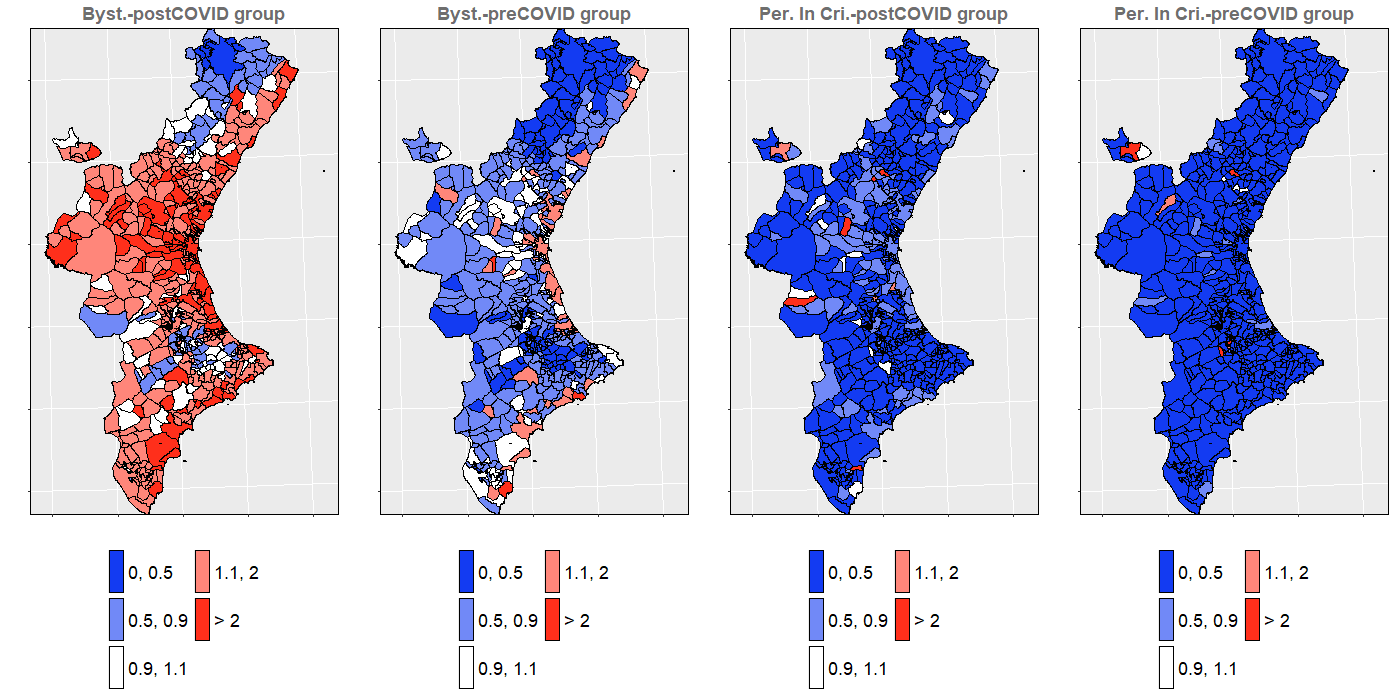}
    \caption{Relative Risks adjusted for the best model of Caller VS Period in suicide-related emergency calls.}
    \label{fig:4}
\end{figure}

Adjusted relative risks are represented in Figure \ref{fig:4}. As expected by the number of counts, the bystander postCOVID-19 group presents high risk values in most of the region, with the exception of a cluster of areas in the inland north and the inland south, which coincides with the spatial effects observed. Bystander preCOVID-19 presents several moderately high risk areas in the coastline, but the rest of the region has low risk. Finally, both person in crisis pre and postCOVID-19 have low risk in all the region, with the exception of several individual municipalities. 

\subsection{Gender VS Caller}

\hspace{0.2cm} Lastly, we will analyze whether the spatial risk pattern of male and females in crisis varies depending on whether the person making the call is either a bystander or a person in crisis. In this case, it is assumed that the underlying risk population is equal for all groups, given that an individual may be both a bystander and a person in crisis, and that the female and male populations are balanced. \\

The best 5 models in terms of DIC are presented in Table \ref{table4}. For this combination of factors, all of them correspond to $M_6$. In particular, the best model uses male and bystander as basis levels, with gender being the first factor and caller the second factor. 

\begin{table}[!htbp]
\tbl{Top 5 Gender VS Caller models in terms of DIC.}
        {\begin{tabular}{lcccc} \toprule
         \textbf{Model} & \textbf{Combination} & \textbf{DICs} & \textbf{WAICs} & \textbf{CPU (sec)} \\
        \midrule
        $M_6$ & Male * bystander & 8658.4 & 8684.4 &  9.71 \\
        $M_6$ & Female * bystander & 8662.7 & 8678.3 & 7.65 \\
        $M_6$ & bystander * Female & 8686.9 & 8711.1 & 7.84 \\
        $M_6$ & bystander * Male & 8720.3 & 8752 & 8.11 \\
        $M_6$ & person in crisis * Female & 8733.3 & 8727.2 & 9.37 \\
        \bottomrule
        \end{tabular}}
    \label{table4}
\end{table}

\begin{figure}[!htbp]
    \centering
    \includegraphics[scale=0.28]{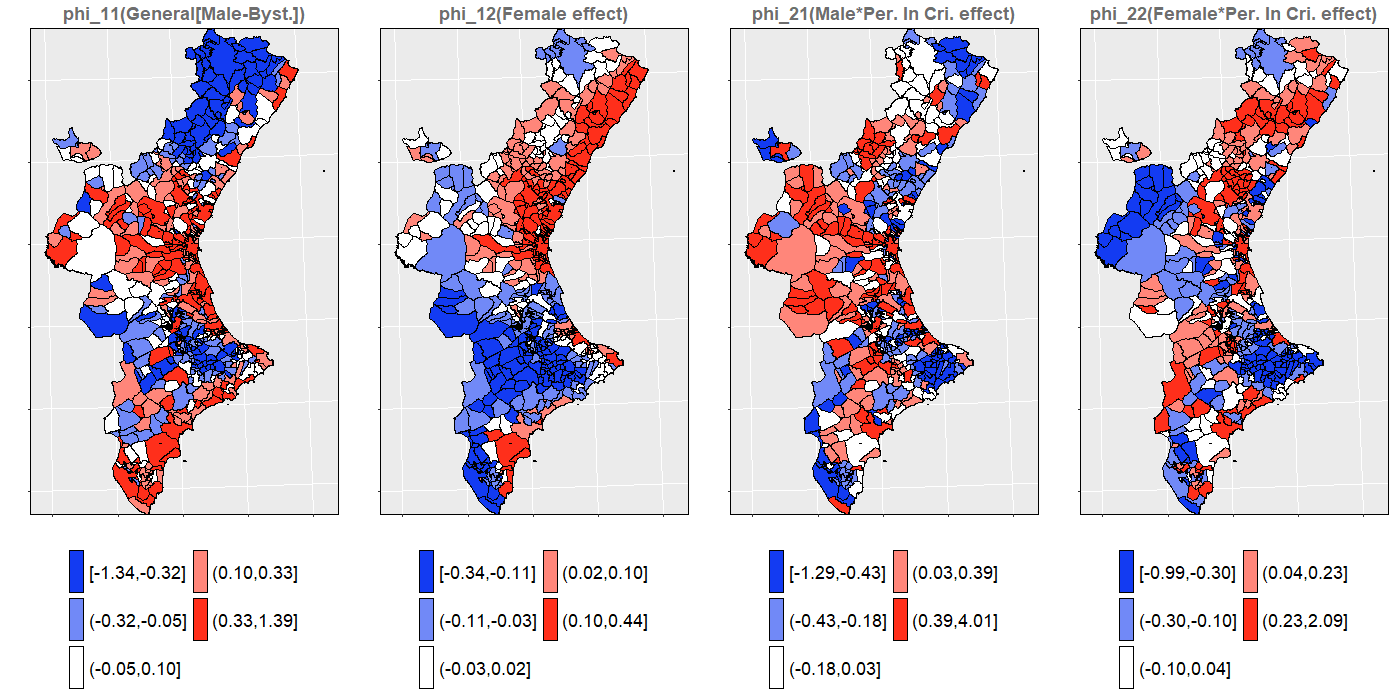}
    \caption{Spatial Effects of the best model for Gender VS Caller in suicide-related emergency calls.}
    \label{fig:5}
\end{figure}

Figure \ref{fig:5} represents the spatial effects adjusted for the best model. For the overall shared effect, there is a low risk zone in the inland north, and a high risk zone in the south of the region. Moreover, the municipalities in the middle of the region exhibit greater values as well. In this case, this effect ranges between -1.34 and 1.39. \\

The female effect oscillates between -0.34 and 0.44, slightly modifying the overall shared effect obtained with male bystander as reference group. This effect presents a greater risk in the north, especially in the coastline, and a lower risk in the south, especially inland. \\ 

The spatial effect for the male person in crisis interaction reflects a big cluster of high risk municipalities in the mid inland of the region, ranging between -1.29 and 4.01. Finally, the female person in crisis effect ranges between -0.99 and 2.09, showing some high risk groups of municipalities in the inland north and south, with a cluster of lower risk municipalities in the central inland part of the region. \\

\begin{figure}[!htbp]
    \centering
    \includegraphics[scale=0.28]{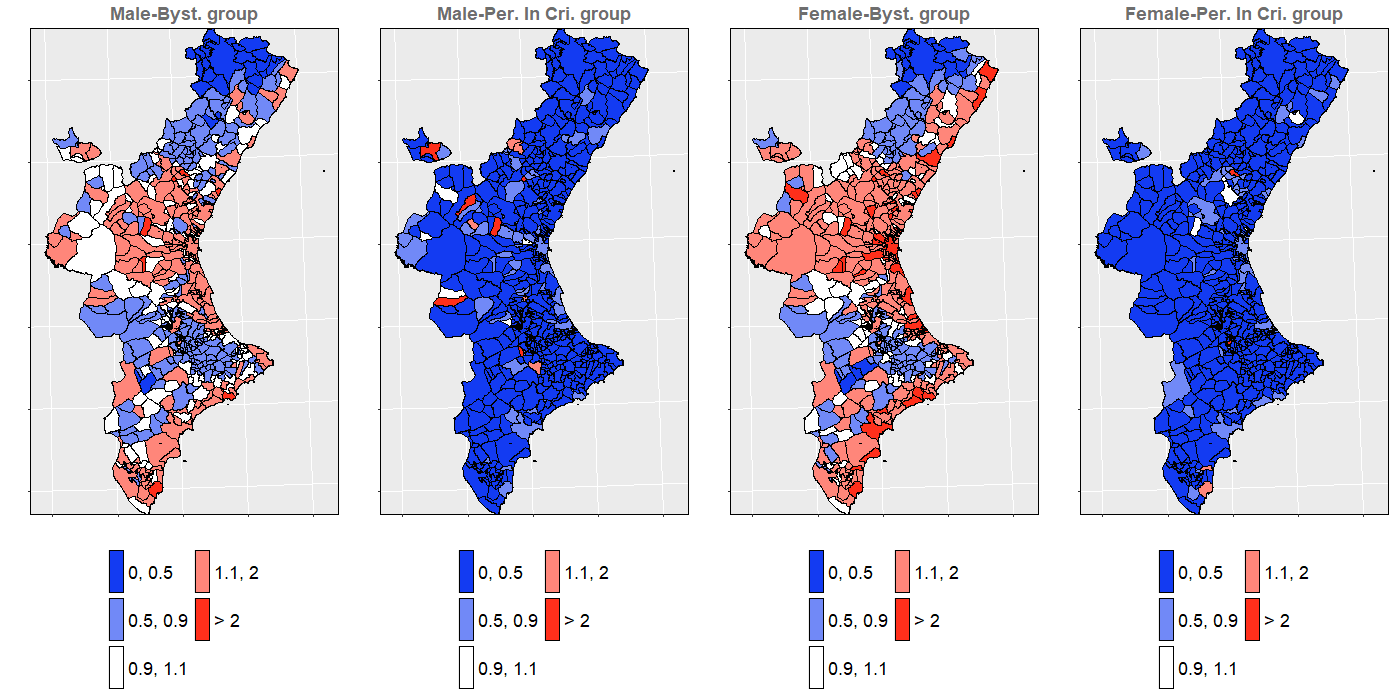}
    \caption{Relative Risks adjusted for the best model of Gender VS Caller in suicide-related emergency calls.}
    \label{fig:6}
\end{figure}

Lastly, Figure \ref{fig:6} represents the adjusted relative risks for the best model. In this case, male and female bystanders present similar patterns. Both groups exhibit clusters of high risk areas in the inland central part of the region, as well as the whole coastline. Female and male person in crisis present most of the municipalities as low risk, which is to be expected considering the number of counts for each group. 

\section{Discussion}

\hspace{0.2cm} Taking advantage of our formulation, we have been able to detect clearly defined spatial risks patterns for the different subgroups studied, with an extremely reasonable computation cost. The results presented analyzing suicide-related emergency calls provide a complementing view for the existing literature of this major social problem. Most of the studies published so far have relied on either mortality data or suicide-related health records \citep{Congdon11, Bridge2018AgeRelatedRD, Helbich2017SpatiotemporalSR}, with some studies using suicide-related emergency calls \citep{Lersch2020COVID19AM, Marco18, Marco2024TheSD}. By the time of this publication, this is the first article where several emergency calls from different subgroups of the population have been modeled and compared using multivariate disease mapping. The potential causal factors underlying the observed patterns remain to be identified. However, it is evident that different population strata are exposed to different risk factors, which are themselves spatially distributed. \\

Further research should consider the addition of sociodemographic covariates, with the potential for investigation into whether the regression coefficients remain consistent across all of the considered subgroups. Nevertheless, this would create new identification problems, with spatial confounding occurring between the regression coefficients and the different shared spatial effects included \citep{Urdangarin2022EvaluatingRM}. \\

Furthermore, the number of groups and categories that can be analyzed using the proposed formulation is a notable limitation. Future research could be conducted with the aim of increasing the number of groups and categories per group, or both. Nevertheless, it is evident that any of these options would significantly increase the number of models that must be considered. Specifically, the addition of groups would introduce higher-order interactions, which could be challenging to identify if the dataset is not highly informative. The increase in categories per group would necessitate testing whether all categories are distinct or not, both within and between groups. \\

Another possible expansion of this formulation would be including a temporal dimension, leading to the development of multivariate spatio-temporal formulations. In this case, it is necessary to consider different specifications, given that both space and time could interact with the categories of the different groups included. Furthermore, spatio-temporal models typically include an interaction term between space and time. Despite the existence of four distinct types, as outlined by \citep{KnorrHeld2000BayesianMO}, the prevalent approach involves utilizing the Type IV interaction. This interaction enables each spatio-temporal unit to influence the overall spatial and temporal effects, resulting in a highly intricate precision matrix, which our formulation is trying to avoid in the first place. \\

In conclusion, this publication presents a new strategy for modelling multidimensional data, whereby several subgroups of the population may exhibit disparate spatial patterns. The principal advantages are a straightforward design, which facilitates the communication of results to a broader audience of social psychology researchers, and the low computational cost. The code of the function, as well as the code for the figures, can be found in \url{https://github.com/VdaK1NG/SRP-ANOVA}. 

\section*{Acknowledgments}

We wish to thank the 112 Valencian Community Emergency Service (Valencia Agency of Security and Response to Emergencies) for providing the dataset.

\section*{Disclosure statement}

The authors of this article declare no conflict of interest.

\section*{Funding}

This study was funded by the Social Observatory La Caixa Foundation (LCF/PR/SR21/52560010). ALQ and FPP thank support by the grant PID2022-136455NB-I00, funded by Ministerio de Ciencia, Innovación y Universidades of Spain (MCIN/AEI/10.13039/501100011033/FEDER, UE) and the European Regional Development Fund.

\end{document}